\documentstyle[12pt]{article}
\begin{document}
\pagestyle{empty}
\setlength{\oddsidemargin}{0.5cm}
\setlength{\evensidemargin}{0.5cm}
\setlength{\footskip}{2.0cm}
\newcommand{\be}{\begin{eqnarray}}
\newcommand{\en}{\end{eqnarray}}
\newcommand{\csl}{{\sl c}}
\newcommand{\ssl}{{\sl s}}
\newcommand{\dr}{{\mit\Delta}r}
\newcommand{\drmt}{{\mit\Delta}r\lbrack m_t \rbrack}
\newcommand{\Born}{\lbrack{\rm Born}\rbrack}
\def\R#1{$\lbrack #1 \rbrack$}

\vspace*{-2cm}\noindent
\hspace*{10.6cm} TOKUSHIMA 94-01 \\
\hspace*{10.6cm} January 1994 \\
\hspace*{10.6cm} (hep-ph/9401340) \\

\vspace*{1.85cm}

\centerline{\large{\bf Non-trivial Test of Electroweak
Quantum Effects}}

\vspace*{1.8cm}

\renewcommand{\thefootnote}{*)}
\centerline{\sc \phantom{*)}Zenr\=o HIOKI\footnote{E-mail
(BITNET) address: A52071@JPNKUDPC or HIOKI@JPNYITP}}

\vspace*{1.8cm}

\centerline{\sl Institute of Theoretical Physics,~
University of Tokushima}

\vskip 0.3cm
\centerline{\sl Tokushima 770,~JAPAN}

\vspace*{3cm}

\centerline{ABSTRACT}

\vspace*{0.4cm}
\baselineskip=20pt plus 0.1pt minus 0.1pt
Based on recent $W$-mass measurements, the electroweak theory
is tested at non-trivial quantum correction level, i.e., beyond
the Born approximation with $\alpha(M_Z)$ instead of $\alpha$.
We can conclude that some non-Born type corrections must exist
at more than 92 \% confidence level, and the non-decoupling
top-quark corrections are required at 97 \% confidence level.

\vfill
\newpage
\pagestyle{plain}
\renewcommand{\thefootnote}{\sharp\arabic{footnote}}
\setcounter{footnote}{0}
\baselineskip=21.0pt plus 0.2pt minus 0.1pt

Electroweak precision analyses have been performed extensively
ever since high-energy experiments at $M_{W,Z}$ scale started
at CERN, FNAL and SLAC. In particular, quite lots of precise
data on the $Z$ boson from LEP have made it possible to test
the standard electroweak theory with considerable
accuracy.\footnote{There are a lot of papers on this subject.
    I only cite \R{1} among the latest articles here (see
    also \R{2,3} and the references cited therein).}\ 
Thereby, many particle physicists now believe that this theory
(plus QCD) describes correctly phenomena below $O(10^2)$ GeV.

Recently, however, Novikov et al. claimed that the Born
approximation based on $\alpha(M_Z)$ instead of
$\alpha$(=1/137.036) reproduces all electroweak precision
measurements within the $1\sigma$ accuracy \R{3}. This
means that the electroweak theory has not yet been tested at
``non-trivial" level (although I never think testing the $\alpha
(M_Z)$ effects to be trivial). Concerning this problem, Sirlin
stressed that such a non-trivial test is possible through
low-energy hadron physics \R{4}. In fact, his conclusion
is that there is very strong evidence for non-Born effects in
the analysis of the unitarity of the Kobayashi-Maskawa mixing
matrix. He also pointed out that more precise measurements of
$M_W$ and the on-resonance asymmetries are crucial for
high-energy tests.

In this short note, I will study the same issue based on the
recent $W$-mass determination by CDF combined with UA2 data
\R{5}:
\be
M_W^{exp}=80.30\pm 0.20~{\rm GeV}. \label{eqaa}
\en
More concretely, I will examine whether the Born approximation
still works or not, and then focus on the top-quark contribution
which does not decouple, i.e., becomes larger and larger as
$m_t$ increases. It is very significant to test it because the
existence of such effects is a characteristic feature of theories
in which particle masses are produced through spontaneous
symmetry breakdown plus large Yukawa couplings.

First, it is quite easy to see if taking only $\alpha(M_Z)$ into
account is still a good approximation. The $W$-mass is calculated
within this approximation as
\be
M_W^2\Born ={1\over 2}M_Z^2
\biggl\{ 1+
\sqrt{\smash{1-{{2\sqrt{2}\pi\alpha(M_Z)}\over{M_Z^2 G_F}}}
\vphantom{A^2\over A}
}~\biggr\}, \label{eqbb}
\en
where $\alpha(M_Z)=1/(128.87\pm 0.12)$ \R{6}. For the
present data $M_Z^{exp}=91.187\pm 0.007$ GeV \R{7} (and
$G_F$=$1.16639\times 10^{-5}$ GeV$^{-2}$), this equation gives
\be
M_W\Born =79.955\pm 0.018~{\rm GeV}, 
\en
which leads to
\be
M_W^{exp}-M_W\Born~=~0.34\pm 0.20~{\rm GeV}.
\en
Since what we want to know here is whether the left-hand side of
it is really positive and not zero, it is a
one-sided test. Therefore we can conclude that the Born
approximation cannot reproduce $M_W^{exp}$ at more than 95 \%
C.L.(confidence level). In \R{5} is also given another
average value
\be
M_W^{exp}=80.21\pm 0.18\ {\rm GeV},
\en
which comes from the above data plus the one by D0: $M_W^{exp}
=79.86\pm 0.40$ GeV. If we use this, the confidence level of our
conclusion becomes about 92 \%. Anyway, these results indicate
that there must be some non-Born type corrections.

The other test that I wish to do here is on the non-decoupling
top-quark effects. Indirect constraints have been derived on
$m_t$ through these correction terms \R{1}. This is deeply
related to the subject under consideration, but not complete as
a test of such corrections: The fact that $m_t$ can be evaluated
through those non-decoupling terms does not automatically mean
that quantum effects including those terms must exist. What
Novikov et al. did \R{3} shows it, indeed. Moreover, the
Higgs mass $m_{\phi}$ brings inevitably some uncertainties into
those usual analyses since we only know $m_{\phi}\ >$ 62.5 GeV
\R{8} at present. Towards unambiguous tests, I proposed a
simple procedure in Ref.\R{9}. I apply it to the present
data here.

Let me briefly summarize my previous work. The tool is the
well-known $M_W$-$M_Z$ relation (see \R{2,10} as review
articles). Except for the higher order $m_t^2$ contributions,
this relation is given in terms of $\alpha$, $G_F$, $M_Z$ and
$\dr$ (the one-loop corrections to the $\mu$-decay amplitude) as
\be
M_W^2={1\over 2}M_Z^2
\biggl\{ 1+
\sqrt{\smash{1-{{2\sqrt{2}\pi\alpha}\over{M_Z^2 G_F (1-\dr)}}}
\vphantom{A^2\over A}
}~\biggr\}. \label{eqcc}
\en
The explicit expression of $\dr$ is given, e.g., in \R{10}.
The non-decoupling top contribution to $\dr$ is
\be
\drmt=-{\alpha\over{16\pi\ssl_W^2}}
\biggl\{ {3\over{\ssl_W^2 M_Z^2}}m_t^2
+4\biggl({\csl_W^2\over\ssl_W^2}-{1\over 3}
-{{3m_b^2}\over{\ssl_W^2 M_Z^2}}\biggr)
\ln\Bigl({m_t\over M_Z}\Bigr)\biggr\}.  \label{eqdd}
\en

\vskip -0.85cm
$$
(\ \csl_W^{\phantom 2} \equiv M_W/M_Z~~{\rm and}
{}~~\ssl_W^2 = 1 - \csl_W^2)
$$
What I proposed is to study what will happen if $\drmt$ would
not exist, i.e., to compute the $W$-mass by using the following
$\dr'$ instead of $\dr$ in Eq.(\ref{eqcc}):
\be
\dr'\equiv \dr-\drmt.
\en
The resultant $W$-mass is denoted as $M_W'$. The important point
is to subtract not only $m_t^2$ term but also $\ln(m_t/M_Z)$
term, though the latter produces only very small effects unless
$m_t$ is extremely large. $\dr'$ still includes $m_t$ dependent
terms, but no longer diverges for $m_t\to +\infty$ thanks to
this subtraction. I found that $M_W'$ takes the maximum for the
largest $m_t$ (i.e., $m_t\to+\infty$) and for the smallest
$m_{\phi}$ (i.e., $m_{\phi}=$62.5 GeV). The accompanying
uncertainty was estimated at most to be about 0.03 GeV.
Therefore,
\be
M_W'\ <\ 79.862\ (\pm 0.030)\ \ {\rm GeV}
\en
holds for any experimentally allowed values of $m_t$ and
$m_{\phi}$.

Let us compare this inequality with $M_W^{exp}=80.30\pm 0.20$
GeV (CDF+UA2). Then, we have
\be
M_W^{exp}-M_W'\ >\ 0.44\pm 0.20\ {\rm GeV},
\en
which shows that $M_W'$ is in disagreement with $M_W^{exp}$ at
almost 99 \% C.L..\footnote{We should also note that the tree
    prediction for the $W$ mass (Eq.(\ref{eqcc}) with $\dr=0$)
    is $M_W^{(0)}=$80.938$\pm$0.009 GeV, which deviates from
    $M_W^{exp}$ at more than 99.9 \% C.L..}\ 
If we use $M_W^{exp}=80.21\pm 0.18$ GeV (CDF+UA2+D0), the above
statement is at about 97 \% C.L.. On the other hand, how about
the $W$ mass calculated with the whole corrections? As an
example, I show it (expressed as $M_W$) for $m_t=$150 GeV and
$m_{\phi}=$100 GeV:
\be
M_W=80.26\pm 0.03\ {\rm GeV}, 
\en
where I have included $m_t^2$ term resummation \R{11} plus
QCD corrections (with $\alpha_{\rm QCD}(M_Z^2)$=0.12) to the
top-quark loop \R{12}. The agreement with the data is quite
good, which is already known very well.

This means that 1) the electroweak theory cannot reproduce
$M_W^{exp}$ \underline{whatever} \underline{values $m_t$ and
$m_{\phi}$ take} if the non-decoupling top-quark corrections
$\drmt$ would not exist, and 2) the theory with $\drmt$ works
very well for $m_t\sim$ 150 GeV, which is consistent with the
present bound: $m_t>$113 GeV \R{13}. Combining them, we are
lead to an interesting phenomenological indication that the
latest experimental data of $M_{W,Z}$ demand,
\underline{independent of $m_{\phi}$}, the existence of the
non-decoupling top-quark corrections. It is a very important test
of the electroweak theory as a renormalizable quantum field
theory with spontaneous symmetry breakdown.

In summary, I have carried out here analyses on the electroweak
quantum corrections beyond the Born approximation with $\alpha
(M_Z)$, and also on the non-decoupling top corrections. We can
thereby conclude that non-Born type corrections are demanded by
the recent data on $M_W$ (CDF+UA2) at more than 95 \% C.L.
(92 \% C.L. if we use the data from CDF, UA2 and D0), and
non-decoupling $m_t$ contribution is required at almost 99 \%
C.L. (97 \% C.L.). This is a clean test of the electroweak
quantum effects which has the least dependence on hadronic
contributions.

\vfill
\newpage
\centerline{ACKNOWLEDGEMENTS}

\vspace*{0.3cm}
I would like to thank W. Hollik and S. Matsumoto for
correspondences.

\vskip 0.8cm
\def\APP#1#2#3{{\sl Acta Phys. Pol.}\ {\bf #1}\ (#2),\ #3}
\def\MPL#1#2#3{{\sl Mod. Phys. Letters}\ {\bf #1}\ (#2),\ #3}
\def\NP#1#2#3{{\sl Nucl. Phys.}\ {\bf #1}\ (#2),\ #3}
\def\PL#1#2#3{{\sl Phys. Letters}\ {\bf #1}\ (#2),\ #3}
\def\PR#1#2#3{{\sl Phys. Rev.}\ {\bf #1}\ (#2),\ #3}
\def\PRL#1#2#3{{\sl Phys. Rev. Letters}\ {\bf #1}\ (#2),\ #3}
\def\PTP#1#2#3{{\sl Prog. Theor. Phys.}\ {\bf #1}\ (#2),\ #3}
\def\ZP#1#2#3{{\sl Zeit. f\"ur Phys.}\ {\bf #1}\ (#2),\ #3}
\centerline{REFERENCES}
\begin{itemize}
\item:\R{1}! J. Ellis, G. L. Fogli and E. Lisi,
\PL{B318}{1993}{148};\\
G. Altarelli, Preprint CERN-TH.7045/93.
\item:\R{2}! S. Fanchiotti, B. Kniehl and A. Sirlin,
\PR{D48}{1993}{307};\\
W. Hollik, Preprint MPI-Ph/93-21.
\item:\R{3}! V. A. Novikov, L. B. Okun and M. I. Vysotsky,
\MPL{A8}{1993}{2529}.
\item:\R{4}! A. Sirlin, Preprint NYU-TH-93/11/01.
\item:\R{5}! D. Saltzberg, Preprint FERMILAB-Conf-93/355-E
(to appear in: {\it Proceedings of 9th Topical Workshop
on Proton-Antiproton Collider Physics}, Tsukuba, Japan,
October 18 - 22, 1993).
\item:\R{6}! F. Jegerlehner, in: {\it Proceedings of the
1990 Theoretical Advanced Study Institute in Elementary
Particle Physics}, ed. by P. Langacker and M. Cveti\u c
(World Scientific, Singapore, 1991), p. 476.
\item:\R{7}! R. Tanaka, in: {\it Proceedings of the XXVI
International Conference on High Energy Physics}, Dallas,
Texas, 1992, ed. by J. Sanborn, AIP Conf. Proc. No. 272
(AIP, New York, 1993).
\item:\R{8}! V. Innocente, in: {\it Proceedings of the
XXVIII Rencontres de Moriond}, Les Arcs 1993 (to appear).
\item:\R{9}! Z. Hioki, \PR{D45}{1992}{1814}.
\item:\R{10}! Z. Hioki, \MPL{A6}{1991}{2129}.
\item:\R{11}! M. Consoli, W. Hollik and F. Jegerlehner,
\PL{B227}{1989}{167};\\
R. Barbieri, M. Beccaria, P. Ciafaloni, G. Curci and A. Vicer\'e,
\NP{B409}{1993}{105}.
\item:\R{12}! F. Halzen and B. A. Kniehl, \NP{B353}{1991}{567}.
\item:\R{13}! J. M. Benlloch, Preprint FERMILAB-Conf-93/329-E.
\end{itemize}
\end{document}